\documentstyle[twocolumn]{article}
\begin{document}
\pagestyle{empty}


\twocolumn[
\vspace*{1.5cm}

\noindent {\large Radiative decays of $D$ mesons}\\ \ \\
S. Fajfer$^a$, S. Prelovsek$^a$ and P. Singer$^b$\\ \ \\
$^a$ J. Stefan Institute, Jamova 39, P.O. Box 300, 1001 Ljubljana, Slovenia\\ \ \\
$^b$ Department of Physics, Technion---Israel Institute of Technology, Haifa 32000, Israel\\ \ \\ \ \\

The short-distance contribution due to the $c\rightarrow u\gamma$ penguin as 
well as long-distance contributions are considered in an analysis of 
$D\rightarrow V\gamma$ decays, the latter being dominant. The matrix elements
for nine $D\rightarrow V\gamma$ transitions are calculated using a hybrid
model which combines heavy quark effective theory and the chiral
Lagrangian. We present the expected range of the branching ratios for these
decays, the most frequent ones $D^0\rightarrow \bar{K}^{*0}\gamma$ and
$D^+_s \rightarrow \rho^+\gamma$, being a few times $10^{-4}$.\\ \ \\ \ \\]

\noindent{\bf 1. INTRODUCTION AND THE {\mbox {\boldmath $Q\rightarrow q\gamma$}} 

~TRANSITION}\\

Radiative flavour changing quark transitions $Q\rightarrow q\gamma$, occur
at the one loop level in the electroweak Standard Model and would show up at
four different scales $s\rightarrow d\gamma$, $c\rightarrow u\gamma$,
$b\rightarrow s(d)\gamma$ and $t\rightarrow c(u)\gamma$. These transitions
are of intensive theoretical and experimental  interest, especially since
it has been point out [1] that the QCD enhancement of $b\rightarrow s\gamma$ brings
it into the realm of observability [2]. Here, we are concerned with the
$c\rightarrow u\gamma$ transition, which can be expected to induce hadronic
decays like $D\rightarrow V\gamma$.

The amplitude $A^{Q\rightarrow q\gamma}$ for the transition of a heavy
quark $Q$ to a light quark $q$ and an on-shell photon is given by [3]
\begin{eqnarray}
&&\hspace*{-0.75cm}A^{Q\rightarrow q\gamma} = \frac{eG_F}{4\pi^2\sqrt{2}} \sum_\lambda
V_{\lambda Q} V^*_{\lambda q} \bar{u} (q) F_{2,\lambda}(k^2)
i\sigma_{\mu\nu}k^\nu \nonumber \\
&&\hspace{0.75cm} \left(m_Q\frac{1+\gamma_5}{2	} + m_q \frac{1-\gamma_5}{2}\right)
u(Q)\epsilon^{\mu\dagger}
\end{eqnarray}
where in our case $F_2=\Sigma_\lambda V_{\lambda c} V^*_{\lambda u}
F_{2,\lambda}$ and the summation is over $\lambda = d,s,b$. $F_2$ is
calculable in the electroweak model [3]; furthermore, the inclusion of\
QCD corrections leads to a strong enhancement of its value [4,5].
Despite this increase, the short distance $c\rightarrow u\gamma$ 
amplitude leads to an inclusive branching ratio of $10^{-8}$ only [5];
accordingly, one expects for the exclusive transitions $D\rightarrow V\gamma$ a 
branching ratio of about $10^{-9}$ from this short-distance contribution.

The long-distance (LD) contribution to $c\rightarrow u\gamma$ has also been
estimated recently [6]. Considering the process $c\rightarrow u +
\bar{f}_if_i$ with the quark pairs $\bar{f}_if_i$ hadronizing into vector
mesons,  one derives the amplitudes for the $c\rightarrow uV_i$
processes ($V_i=\rho,\omega,\varphi)$ using the QCD corrected nonleptonic
weak Lagrangian. Taking then the transverse part of the $c\rightarrow uV_i$
amplitude, with gauge invariance and vector meson dominance one arrives at 
the amplitude [6]

\begin{eqnarray}
&&\hspace{-0.80cm} ^{LD}A(c\rightarrow u\gamma)=\frac{G_F}{\sqrt{2}} a_2
(m^2_c) V_{ud} V^*_{cd}\frac{C'_{\rm VMD}}{m_c}\nonumber \\
&&\hspace{1.75cm} \left[\bar{u}\sigma^{\mu\nu} (1+\gamma_5)c\right] k_\nu\epsilon^*_\mu
\end{eqnarray}
where $a_2$ is a Wilson coefficient and $C'_{\rm VMD}$ is
given by
\begin{eqnarray}
C'_{\rm VMD}=-\frac{1}{2} \frac{g^2_\rho(0)}{m_\rho^2} + \frac{1}{6}
\frac{g^2_\omega(0)}{m_\omega^2} + 
\frac{1}{3} \frac{g^2_\varphi(0)}{m_\varphi^2} \ .
\end{eqnarray}
Using the vector-photon couplings measured in leptonic decays, and assuming
$g_{V_i}(0)\simeq g_{V_i}(m^2_{V_i})$ one finds
[7] a strong cancellation in (3) due to the GIM mechanism and SU(3) symmetry,
leading again to small branching ratios, e.g. about $10^{-8}$ for 
$D^0\rightarrow \rho^0\gamma$ [6], only slightly larger than the SD
contribution.\\

\noindent{\bf 2. A MODEL FOR {\mbox{\boldmath $D\rightarrow V\gamma$}}

~TRANSITIONS}\\

The results of the previous section indicate the need for a detailed
treatment of the LD contributions to the $D\rightarrow V\gamma$ transitions;
indeed, several approaches were put forward using a pole model [4], a quark
model [8] and QCD sum rules [9].
In the
present treatment, we aim for a more comprehensive and systematic treatment
for these decays, employing an effective hybrid Lagrangian [10] which
combines two approximate symmetries of QCD, the infinite heavy quark ($Q=c)$
mass limit and the chiral limit for light quarks.

We assume [6,11] the radiative decays to originate from the nonleptonic
weak transition $D\rightarrow VV_0$ [4,12] followed by the conversion
$V_0\rightarrow \gamma$ via vector meson dominance. In addition, there are
transitions due to direct photon emission from the initial $D$ state.
The effective nonleptonic Lagrangian we use is given by
\begin{eqnarray}
{\cal L}_{LD} &=& - \frac{G_F}{\sqrt{2}} V_{uq_i} V^*_{cq_j}
[a_1(\bar{u}q_i) ^\mu(\bar{q}_jc)_\mu + \nonumber \\
&& a_2(\bar{u}c)_\mu(\bar{q}_j q_i)^\mu]
\end{eqnarray}
and in order to evaluate the matrix elements of the product of two
currents we use systematically [11] the factorization approximation.
The quark bilinears in (4) are treated as interpolating fields
for the appropriate mesons, the relevant hadronic degrees of
freedom being the charm pseudoscalar ($D)$ and vector mesons $(D^*)$
and the light pseudoscalar ($P$) and vector mesons ($V)$. In the 
factorization approach we use, the $D\rightarrow VV_o$ amplitude is schematically
approximated as
\begin{eqnarray}
&& \hspace*{-1cm} \langle VV_0|(\bar{q}_iq_j)^\mu(\bar{q}_kc)_\mu|D\rangle=\nonumber \\
&&\hspace{1cm}\langle V|(\bar{q}_iq_j)^\mu|0\rangle
\langle V_0|(\bar{q}_kc)_\mu |D\rangle\nonumber \\
&& \hspace{1cm}+ \langle V_0|(\bar{q}_iq_j)^\mu|0\rangle\langle V|(\bar{q}_kc)_\mu|D\rangle\nonumber \\
&& \hspace{1cm} + \langle VV_0|(\bar{q}_iq_j)^\mu|0\rangle
\langle 0|(\bar{q}_kc)_\mu| D\rangle 
\end{eqnarray}
There are three terms in the factorized $D\rightarrow VV_0$ amplitude,
referring to an annihilation part, a $V_0$-spectator part and a $V$-spectator
part, each of them involving several diagrams.
For the QCD-induced constants $a_1,a_2$ we take $a_1=1.26$, $a_2=-0.55$
as determined from an extensive application of ($4$) to nonleptonic $D$
decays [13].

The general gauge invariant amplitude for the decay $D(p)\rightarrow
V(p_V)\gamma(k)$ is
\begin{eqnarray}
&&\hspace*{-1cm} A(D\rightarrow V+\gamma)=\frac{eG_F}{\sqrt{2}} V_{uq_i}
V_{cq_i}^*\nonumber \\
&&\hspace*{1cm}\left\{ \epsilon_{\mu\nu\alpha\beta}k^\mu\epsilon^{*\nu}_{(\gamma)}
p^\alpha \epsilon_{(V)}^{*\beta}A_{PC}\right.\nonumber \\
&&\hspace*{1cm} + i \left[\left(\epsilon^*_{(V)}\cdot k\right)
\left(\epsilon^*_\gamma \cdot p_{(V)}\right)\nonumber \right.\\
&&\hspace*{1cm} \left. \left. -\left(p_{(V)}\cdot k\right)
\left(\epsilon^*_{(V)}\epsilon^*_{(\gamma)}\right)\right]\right\} 
A_{PV} \ . 
\end{eqnarray}
We have classified all diagrams contributing to $A_{PC}$, $A_{PV}$,
and the explicit expressions are
given in references [6] and [11].

The calculation requires the $\langle V|\left(V_\mu-A_\mu\right)D\rangle$
matrix element, which has one vector $V(q^2)$ and four axial-vector $A_i(q^2)$
form factors. Actually, gauge invariance and finiteness reduce the number
of independent form factors to three. For these we take a pole dominance
behaviour, with known values of pole masses. Information on $V(0)$,
$A_1(0)$, $A_2(0)$ we extract from the semileptonic decays 
$D^+\rightarrow \bar{K}^{*0} \ell\nu_e$ and $D_s^\dagger\rightarrow
\varphi \ell\nu_e$.\\

\noindent{\bf 3. RESULTS}\\

The calculation of the decay amplitudes involves various constants of the
hybrid Lagrangian. Their imprecise knowledge is the first source of
uncertainty. In addition, the amplitudes contain several components
with unknown relative phases, which is yet another source of uncertainty.
This allows us to give only an expected range, for each of the calculated
transitions. In the Table below, we give the predicted range of the 
calculated values for the branching ratios.  The first two decays are
Cabibbo-allowed, the next five are Cabibbo-forbidden and the last two
are doubly forbidden.  To give an indication, the photon energy in the
first two decays is 717 and 834 MeV respectively.
As the table shows, a few of these decays are in a range which should become experimentally
feasible soon.
Their detection will
contribute decisively to the understanding of long-distance dynamics
in D decays.

  On the other hand, we are still left with the question
of how to detect directly the $c\rightarrow u\gamma$ transition. One
possibility is related to accidental cancellations between the various
other contributions to the amplitudes of Cabibbo-forbidden decays [6], which is
probably unlikely. The alternative would be to search for weak radiative
transitions among baryonic states containing several $c$ quarks,
like $\Xi^{++}_{cc} \rightarrow \Sigma^{++}_c\gamma$ and
$\Omega^{++}_{ccc} \rightarrow \Xi^{++}_{cc}\gamma$. In these cases,
due to the valence structure of the participating baryons, the $c\rightarrow
u\gamma$ process would play a dominant role[14].
\\ \ \\
\noindent Table 1\\
The predicted branching ratios for nine $D\rightarrow V\gamma$ decays
\begin{center}
\begin{tabular}{lc}\hline
$D\rightarrow V\gamma$ Transition & Br Ratio $\times 10^5$[11]\\[0.10cm] \hline
& \\
$D^0\rightarrow\bar{K}^{*0}\gamma$ & 6-36\\[0.10cm]
$D_s^+\rightarrow\rho^+\gamma$ & 20-80\\[0.10cm]
$D^0\rightarrow\rho^0\gamma$ & 0.1-1\\[0.10cm]
$D^0\rightarrow\omega\gamma$ & 0.1-0.9\\[0.10cm]
$D^0\rightarrow\varphi\gamma$  & 0.4-1.9\\[0.10cm]
$D^+\rightarrow\rho^+\gamma$ & 0.4-6.3\\[0.10cm]
$D^+_s\rightarrow {K}^{*+}\gamma$ & 1.2-5.1\\[0.10cm]
$D^+\rightarrow {K}^{*+}\gamma$  & 0.03-0.44\\[0.10cm]
$D^0\rightarrow {K}^{*0}\gamma$  & 0.03-0.2 \\ \hline
\end{tabular}
\end{center}
\vspace{0.50cm}

\noindent{\bf REFERENCES}
\begin{enumerate}
\item S. Bertolini, F. Borzumati, and A. Masiero, Phys.\ Rev.\ Lett.\ 
59 (1987) 180; N.G. Deshpande, P. Lo, J.\ Trampetic, G.Eilam, and P. Singer,
Phys.\ Rev.\ Lett. 59 (1987) 183.\vspace{-0.25cm}

\vfill
~
\pagebreak

\item M.S. Alam et al. (CLEO Collaboration), Phys.\ Rev.\ Lett.\ 74 (1995)
2885.\vspace{-0.25cm}
\item T. Inami and C.S. Lim, Prog.\ Theor.\ Phys.\ 65 (1981) 297.\vspace{-0.25cm}
\item G. Burdman, E. Golowich, J.L. Hewett and S. Pakvasa, Phys.\ Rev.\ D
52 (1995) 6383.\vspace{-0.25cm}
\item C. Greub, T. Hurth, M. Misiak and D. Wyler, Phys.\ Lett. B 382 (1996) 415.\vspace{-0.25cm}
\item S. Fajfer and P. Singer, Phys.\ Rev. D56 (1997) 4302.\vspace{-0.25cm}
\item P. Singer and D.-X. Zhang, Phys.\ Rev.\ D54 (1996) 1225; 
G. Eilam, A. Ioannissian, R.R. Mendel and P. Singer, Phys.\ Rev.\
D53 (1996) 3629.\vspace{-0.25cm}
\item H.-Y. Cheng et al., Phys.\ Rev.\ D51 (1995) 1199.\vspace{-0.25cm}
\item A. Khodjamirian, G. Stall and D. Wyler, Phys.\ Lett.\ B358 (1995) 129.\vspace{-0.25cm}
\item R. Casalbuoni et al., Phys.\ Reports 281 (1997) 145.\vspace{-0.25cm}
\item S. Fajfer, S. Prelovsek and P. Singer, Eur.\ Phys.\ Journal C (in print).\vspace{-0.25cm}
\item B. Bajc, S. Fajfer and R.J. Oakes, Phys.\ Rev.\ D51 (1995) 2230.\vspace{-0.25cm}
\item M. Bauer, B. Stech and M. Wirkel, Z.\ Phys.\ C34 (1987) 103.\vspace{-0.25cm}
\item P. Singer, Nucl.\ Phys.\ B. (Proc.\ Suppl.) 50 (1996) 202.
\end{enumerate}

\end{document}